# Surface states induced weak anti-localization effect in $Bi_{0.85}Sb_{0.15}$ topological single crystal

Yogesh Kumar[1,2] and V.P.S. Awana[1,2*]

[1]*CSIR-National Physical Laboratory, Dr. K. S. Krishnan Marg, New Delhi-110012, India*

[2]*Academy of Scientific and Innovative Research (AcSIR), Ghaziabad 201002, India*

**Abstract**

We report, an experimental evidence of surface states (SS) driven magneto-transport in a $Bi_{0.85}Sb_{0.15}$ single crystal. Detailed high field (up to 12T) and low temperature (down to 2K) magneto-transport measurements are been carried out on the studied $Bi_{0.85}Sb_{0.15}$ single crystal. The phase, composition and Raman modes are studied through X-ray diffraction, Energy dispersive X-ray, and Raman spectroscopy. The obtained crystal shows non-saturating magnetoresistance ($\approx 4250\%$) at 2K and 12T, along with the existence of weak-anti localization (WAL) effect at around zero magnetic field. Further, the Hikami-Larkin-Nagaoka (HLN) analysis is performed to analyse the WAL effect. The prefactor ($\alpha$) and phase coherence length ($L_\varphi$) are deduced at various temperatures, which signified the presence of more than one conduction channel in the studied $Bi_{0.85}Sb_{0.15}$ single crystal. The effect of quantum scattering, bulk contribution from underneath the surface states and defects are been studied by adding various field dependent quadratic, linear and constant terms to the SS driven HLN equation. Various possible scattering mechanism are studied by analysing the temperature dependence of the phase coherence length. Angle dependent magneto-conductivity of the studied $Bi_{0.85}Sb_{0.15}$ single crystal clearly confirmed the surface states dominated transport in present crystal.

*****Corresponding Author**

Dr. V. P. S. Awana:  E-mail: awana@nplindia.org
Ph. +91-11-45609357, Fax-+91-11-45609310
Homepage: awanavps.webs.com

**Introduction**

Topological insulators (TIs) are quantum materials possessing both insulating bulk states underneath and the highly conducting channels at the surface. The occurrence of band inversion due to presence of spin-orbit coupling (SOC) results in highly conducting surface states, which are protected by time reversal symmetry [1,2]. The surface state electrons are less sensitive to non-magnetic impurities and thus have high mobility due to topological protection of metallic surface states [3,4]. These TIs are of much interest for possessing various properties such as monopole magnets, Majorana fermions, superconducting proximity effect and topological quantum computation [5,6]. Bismuth based chalcogenides were confirmed theoretically and experimentally to possess these topological surface states. Experimental techniques such as angle-resolved photoemission spectroscopy (ARPES) and scanning tunnelling microscopy (STM) were used to probe these metallic surface states [7-9]. The semi-metallic phase of pure bismuth can be modified by partial substitution of bismuth by antimony. The added antimony changes the spin-orbit interaction and thus the band structure. A transition from semimetal to insulator phase has been observed, depending upon the concentration of antimony in $Bi_{1-x}Sb_x$, on the other hand the same becomes topological insulator for intermediate concentrations i.e., x = 0.07-0.22 [10,11]. Besides the ARPES and STM, another way to probe the topological surface states is to study magnetic field dependent electrical transport of TIs.

In solids, the quantum interference of electrons undergoes backscattering from scattering centres in time-reversed loops. The constructive or destructive interference between two time-reversed loops leads to localization or delocalization of electrons. In TIs, the presence of π Berry phase supresses the backscattering of electrons due to destructive interference between time-reversed loops and this delocalization is known as WAL effect. When magnetic field is applied to such a system, it destroys the interference and is seen as negative magneto-conductivity in TIs [12-14]. The TIs are by now widely known to exhibit WAL effect in their magneto-transport measurements at low temperature and in and around low magnetic field regime, which is widely studied by Hikami-Larkin-Nagaoka (HLN) formula [15,16]. By analysing the magneto-conductivity of TIs using HLN model, the number of conduction channels contributing in overall transport mechanism, and the phase coherence length of surface carriers can be estimated [16].

Keeping in view, the importance of low temperature and high field magneto-transport of TIs, in this article, we report the same for as grown and well characterized (structure/microstructure/spectroscopic) $Bi_{0.85}Sb_{0.15}$ single crystel, down to 2K and up to 12T field. Detailed magneto-transport measurements were performed on mechanically cleaved thin single crystalline flakes at various temperatures.

Weak anti-localization effect is analysed by Hikami-Larkin-Nagaoka (HLN) equation, and the characteristic parameters i.e., prefactor (α) and phase coherence length ($L_\varphi$) are extracted. Further, terms explaining quantum scattering, cyclotronic MR and bulk contribution are added in HLN formula. The angle dependent MR is also studied by varying the angle between applied field and the direction of current, which demonstrated 2D nature of surface conduction electrons. The temperature dependence of extracted value $L_\varphi$ is further studied by Kohler's rule and $L_\varphi(T)$ equation, which revealed the presence of electron-electron and electron-phonon scattering. In nut shell this is a detailed magneto-transport study on a well characterized $Bi_{0.85}Sb_{0.15}$ single crystal.

**Experimental details**

Solid state reaction route via self-flux method is used to synthesize and grow $Bi_{0.85}Sb_{0.15}$ single crystals. The stoichiometric amount of high purity (99.999%) bismuth and antimony were ground thoroughly in inert atmosphere inside the MBRAUN glove box. The binary mixture then pelletized and sealed into an evacuated quartz tube. The single crystals were grown by following the heat treatment as reported earlier [17]. The quartz tube was heated to 650°C, and kept hold for 8 hours in order to get homogenized mixture. It was then slowly cooled down to 250°C with cooling rate 3°C/hour and kept hold for 95 hours. Again, the tube was cooled down to 245°C with a cooling rate of 2°C/hour and kept hold for 73 hours. Finally, the ampoule was cooled to room temperature with rate of 120°C/hour and the silvery shiny single crystal was obtained. Thin flakes were mechanically cleaved from as grown single crystal for further characterization and measurements. X-ray diffraction (XRD) spectra were recorded on both thin flake and finely crushed powder of as grown crystal using Rigaku MiniFlex II, DESKTOP X-ray diffractometer equipped with Cu-$K_\alpha$ radiation (λ=1.5418 Å). FullProf and VESTA software were used to perform Rietveld refinement and draw unit cell structure, respectively. Scanning electron microscope (SEM) image and Energy dispersive X-ray analysis (EDAX) were performed using Zeiss EVO-50 system. Raman spectra were recorded at room temperature using Renishaw in Via Reflex Raman microscope equipped with 514 nm Laser. Electrical and magneto-transport measurement were performed on thin single crystal flakes in conventional four-probe geometry, using Quantum Design Physical Property Measurement System (QD-PPMS).

**Results and Discussion**

To determine the crystal structure and phase purity of as grown $Bi_{0.85}Sb_{0.15}$ single crystal, powder X-ray diffraction (XRD) spectra were recorded on finely crushed powder. In fig. 1(a), red symbol shows the observed room temperature diffraction spectra from a 2θ range of 20° to 80°, whereas solid black line represents the calculated intensity peaks extracted from Rietveld refinement using FullProf software. The grown crystal belongs to $R\bar{3}m$ (166) space group with rhombohedral crystal structure and all the diffraction peaks are indexed with their respective planes. The quality of goodness of fit is confirmed by the parameter $\chi^2$ which comes out to be 2.33. The lattice parameters obtained from Rietveld refinement are a = b = 4.507(3) Å and c = 11.768(4) Å. The absence of any additional diffraction peak signifies that there is no other phase or impurity present in the grown sample. Further to study the single crystalline nature, XRD spectra were recorded on mechanically cleaved thin flake of as grown $Bi_{0.85}Sb_{0.15}$ sample. As shown in fig. 1(b), the XRD spectra of thin flake reveals that the crystal is grown along the (00$l$) diffraction plane, where $l$ = 3,6,9, which suggests the growth is along the c-axis. In addition to (00$l$) peaks, a very small intensity peak corresponding to (202) diffraction plane is also observed, which may be due to misalignment of plane in crystal flake. The inset of fig. 1(a) represents the unit cell structure of grown $Bi_{0.85}Sb_{0.15}$ single crystal, where red and black solid sphere corresponds to Bi and Sb atoms, respectively. The unit cell is extracted from VESTA software by using the cif (Crystallographic Information File) file, which is generated from Rietveld refinement.

The surface morphology of synthesized crystal was studied through SEM image, which is shown in fig. 2(a). The obtained image depicts the steps like layered structure, which corresponds to layered growth of single crystal. Further, the purity of sample is studied by performing the EDAX spectra and is shown in fig. 2(a). The EDAX spectra confirms the absence of any additional elements and also the concentration of respective elements is near to their initial stoichiometric ratio. The right side image in fig. 2(a) represents the EDAX mapping on sample surface, which signifies the uniform distribution of corresponding elements in entire crystal without any cluster formation or segregation. To study the different vibrational modes, Raman spectra were recorded on thin flakes, using 514 nm Laser with power less than 5 mW for 15 sec to avoid any local heating. The distinct Raman peaks were observed as shown in fig.

2(b). The recorded spectra are deconvoluted into four discrete peaks corresponding to Bi-Bi, Sb-Sb and Bi-Sb bonds. The low frequency modes observed at 71.11 cm$^{-1}$ and 95.47 cm$^{-1}$ correspond respectively to $E_g$ and $A_{1g}$ modes of Bi-Bi bond vibrations, whereas another mode at 115.71 cm$^{-1}$ represents the $E_g$ mode of Sb-Sb bond vibrations. Another high frequency Raman mode observed at 129.41 cm$^{-1}$ represents the Bi-Sb bond vibrations. All the observed Raman modes are in good agreement with previous reports [18-20].

Fig. 3(a) shows the temperature dependent resistivity at different magnetic fields, the standard four-probe geometry was used to measure electrical resistivity and the magnetic field was applied perpendicular to the direction of the current flow. In zero field, the resistivity continuously decreases with decreasing temperature from 300K down to 2K which depicts the metallic behaviour of grown Bi$_{0.85}$Sb$_{0.15}$ single crystal. Further, the effect of magnetic field on electrical transport is studied in low temperature regime of 100K to 2K. It is observed that slope of resistivity changes from positive (metallic behaviour) to negative (semiconducting behaviour) with application of magnetic field. Also slope becomes more negative at low temperatures with increasing the magnetic field strength and the similar behaviour have been observed in various topological materials. According to previous report by D. V. Khveshchenko, this type of behaviour corresponds to opening of gap at band crossing points due to magnetic field and this opening is proportional to applied field strength [21]. Moreover, it is also observed that, at low temperature the electrical resistivity saturates in presence of magnetic field and this result is in accordance with previous reports [22,23]. Further, the temperature dependence of resistance can be explained by the equation $R_{xx} = R_0 + \beta e^{-\left(\frac{\theta}{T}\right)} + \gamma T^2$ as described by P. Mal et. al., [24] and the fitted curve is shown in inset of fig. 3(a). Here, R$_0$ is residual resistance, quadratic term and exponential term represents the electron-electron (e-e) and electron-phonon (e-ph) interactions, respectively. The obtained fitting parameters are R$_0$ = 0.73 mΩ, β = 1.08 mΩ, θ = 171.4 K and γ = 3.28 × 10$^{-9}$ ΩK$^{-2}$ which corresponds to best fit at zero field. Fig. 3(b) shows the plot of magnetoresistance (MR) as a function of applied transverse magnetic field at various temperatures ranging from 250K down to 2K. The resistance is measured by varying the field from -12T to +12T and then symmetrized by using the relation R(H) = [R(H)+R(-H)]/2. The MR percentage is calculated by using the formula MR% = [(R(H)-R(0))/R(0)]×100, where R(H) and R(0) are resistance at applied field and zero field, respectively. A huge MR% of around 4249% is observed at 2K and 12T without any signature of saturation, and this non-saturating MR% behaviour is still observed at 250K but with decreased value of 323% at 12T. The value of MR% at highest measured field is found to be monotonically decreased with increasing the temperature from 2K to 250K. Interestingly, a pronounced dip-like cusp is observed in MR in low-field regime which corresponds to the presence of weak-anti localization (WAL) effect that has reported previously in various topological insulators. The WAL effect in topological materials can arise from spin-orbit coupling (SOC) or topological surface states [25-27].

In order to understand the origin of WAL effect, the angle dependent MR measurements were performed by varying the angle between applied field and direction of current flow. The inset of Fig. 4(a) shows the variation of magneto-conductivity in a field of ±10T at different angles theta at constant temperature 2K. Here, angle $\theta = 0°$ and 90° corresponds to the situations, when field is perpendicular and parallel to current direction, respectively. It is observed that the measured magneto-conductivity curve at different angles overlaps in low field regime, say below 1.5T. Further, the variation of magneto-conductivity as function of perpendicular component (Hcosθ) at 2K is shown in fig. 4(a). It is observed that at low fields (-1.5T to +1.5T), all curves overlap on each other, which indicates that magneto-conductivity at low fields depends only on perpendicular magnetic field component. This corresponds to the contribution from 2D surface electrons in electrical transport. Although, a deviation from universal curve at higher field is also observed in magneto-conductivity curve which may be due to the contribution from bulk channels and the observed data is in accordance with previous reports [25, 28-31]. Another way to understand, whether the conduction mechanism is governed by bulk charge carriers or surface charge carries, is to study the angular dependency of magneto-transport [32,33]. Fig. 4(b) represents the variation in $\rho/\rho_{0°}$ with respect to angle $\theta$ in a magnetic field of 10T at temperature 2K, here $\rho$ and $\rho_{0°}$ are resistivity at respective angle and 0°, respectively. The angle $\theta$ is defined as the angle between normal to sample surface and applied field direction is shown in schematic of fig. 4(b). It is observed that normalized resistivity has oscillating behaviour with maxima in perpendicular field arrangement ($\theta = 0°$ and 180°) and minima in parallel field arrangement ($\theta = 90°$ and 270°). According to literature, the electrical transport of 2D surface carriers will only depend on perpendicular component of applied magnetic field [34,35]. The experimental data is fitted with the functional form $a + b|cos\theta|$ in entire $\theta$ range from 0° to 360° as plotted by solid curve and the extracted values of parameter a and b are 0.4 and 0.64, respectively. The fitted curve is in accordance with experimental data, suggesting dominant contribution from 2D surface charge carriers in the transport mechanism. Albeit, the observed small deviation of fitted |cosθ| curve may indicate the presence of bulk charge carriers in overall transport.

Furthermore, to understand the presence of WAL effect and to study the conduction channels and phase coherence length, the magneto-conductivity is analysed. The conductivity is extracted by inverting the resistivity at respective field and temperature. Fig. 5(a) represents the magneto-conductivity as a function of transverse magnetic field in a range of ±12T at various temperatures from 250K down to 2K. In low field regime, a cusp-like behaviour is observed at all measured temperatures, which becomes sharper with decreasing the temperature. The appearance of negative magneto-conductivity suggests the existence of WAL effect in $Bi_{0.85}Sb_{0.15}$ single crystal [15]. The physical parameters describing the WAL effect can be extracted by using the Hikami-Larkin-Nagaoka (HLN) formula [16]. According the HLN model, for a 2D system the quantum correction to conductivity can be explained by

$$\Delta\sigma(H) = -\frac{\alpha e^2}{\pi h}\left[ln\left(\frac{B_\varphi}{H}\right) - \Psi\left(\frac{1}{2} + \frac{B_\varphi}{H}\right)\right] \quad (1)$$

here, $B_\varphi = \frac{h}{8e\pi L_\varphi^2}$ is characteristic field, $L_\varphi$ is phase coherence length, $\Psi$ is digamma function, h is Plank's constant and e is the electronic charge. The experimentally observed magneto-conductivity is fitted by above equation with free parameters $L_\varphi$ and α. The magnitude and polarity of α describes the number of conduction channels contributing in transport and presence of WAL or weak localization (WL), respectively. For one topological conduction channel, α should take the value -1/2, whereas for two parallel topological surfaces it becomes -1, but experimentally α deviates from these values which may be due to contribution from both bulk and topological surface states [14,31,36,37]. The magneto-conductivity is fitted with HLN equation in low field regime (±1.5T) for transverse field arrangement at different temperatures, and the fitted curve is plotted by solid curve in fig 5(a). At lowest measured temperature 2K, the extracted value of phase coherence length $L_\varphi$ is 57.4nm and pre-factor α is -3.25 which confirms the occurrence of WAL effect and contribution from multiple surface conduction channels. The fitted values of α and $L_\varphi$ at different temperatures are represented in table 1. It is observed that the value of $L_\varphi$ monotonically increases with decreasing the temperature from 250K down to 2K, which indicates the weakening of WAL effect at higher temperatures [30]. Although the occurrence of WAL effect in low field regime is well explained by HLN formula but at higher fields, deviation of HLN fitted curve from experimental data is also observed. This may be due to the contribution from both classical MR and quantum scattering, arising as a consequence of increasing magnetic field strength. The effect of these terms on transport mechanism can be approximated by quadratic dependence on field ($\beta H^2$). Here, the coefficient β comprises of both quantum scattering ($\beta_q$) and cyclotronic ($β_c$) part, further $β_q$ consists of elastic scattering and spin-orbit scattering [38,39]. In addition to field dependent quadratic term explaining scattering contribution, a temperature dependent linear term in field (γH) is also added. According to previous reports on topological insulators, the coefficient of linear term describes the bulk contribution to magneto-conductivity. It is reported that, the transport mechanism is governed by surface carriers at low temperatures but as the temperature increases, there is a bulk contribution in overall transport as well [40,41]. Now, to understand the conduction mechanism at all measured temperatures for entire measured field regime of ±12T, the experimental magneto-conductivity is fitted by equation HLN+$βH^2$+γH+c, here the coefficient β and γ describe the scattering and bulk contributions, respectively and constant c occurs due to any defects in material. In fig. 5(b), the solid curve represents the fitted data in entire field range, which is in good agreement with experimental data. By fitting above equation, the extracted values of all the parameters α, β, γ, c and $L_φ$ are represented in table 2. These values suggest the presence of WAL effect and the transport is governed by surface carriers at low field and temperature, and bulk carriers start contributing with increasing field strength and temperature.

From R-T measurements, it is observed that both e-e and e-ph terms contribute in transport mechanism. So, for deeper understanding of scattering mechanism, MR is investigated as a function of field by Kohler's rule at different temperatures. Kohler's rule states that the change in the resistivity Δρ/$ρ_0$ with respect to applied field H, is described as a function of variable Hτ. Here, Δρ is defined as the difference between resistivity at applied field ($ρ_H$) and resistivity at zero field ($ρ_0$) and τ is defined as an average time between two scattering events of conduction electrons, which is inversely proportional to $ρ_0$. As per Kohler's rule if the Δρ/$ρ_0$ curves with respect to H/$ρ_0$ collapse to a single curve at different temperatures, then the scattering mechanism is governed by single type of charge carriers [42-44]. Fig. 6(a) represents the behaviour of MR% with respect to H/$ρ_0$ at different temperatures. It is observed that the MR curve at different temperatures deviates from a single curve which corresponds to the violation of Kohler's rule in $Bi_{0.85}Sb_{0.15}$ single crystal. The deviation from Kohler's rule confirms the presence of multiple charge carriers at different temperature, and this behaviour is further investigated by studying the effect of temperature on phase coherence length extracted from HLN fitting. According to previous reports, for a two-dimensional system having e-e scattering, the phase coherence length follows the power law dependence on temperature i.e., $L_φ \propto T^{-0.5}$ [31,45]. To understand scattering mechanism, the extracted $L_φ$ from HLN fitting at different temperatures as shown in table 1 are fitted by power law formula as shown in inset of fig. 6(b). From the plot of $(L_\varphi)^{-2}$ vs T, it is observed that the data points deviate from the fitted straight line represented by solid line. This corresponds to the violation of power law behaviour, which suggests that scattering mechanism is not governed by e-e interaction alone and this result is in accordance with previous result from Kohler's rule. Now, in $Bi_{0.85}Sb_{0.15}$ single crystal both the electron-electron and electron-phonon scattering are supposed to emerge in transport mechanism [29]. Hence, the phase length $L_\varphi$ can be defined as a function of temperature in terms of

$$\frac{1}{L_\varphi^2(T)} = \frac{1}{L_\varphi^2(0)} + A_{ee}T^{p_1} + A_{ep}T^{p_2} \qquad (2)$$

here, $L_\varphi(0)$ is phase coherence length at zero temperature, $A_{ee}T^{p_1}$ and $A_{ep}T^{p_2}$ corresponds to e-e and e-ph scattering, respectively. The solid red curve in fig. 6(b) represents the fitting of $L_\varphi(T)$ with respect to temperature, here $L_\varphi(T)$ is extracted from fitting of magneto-conductivity by using equation 1. By considering $p_1 = 1$ and $p_2 = 2$, the obtained values of the coefficient $A_{ee}$ and $A_{ep}$ from fitting are found to be $2.09 \times 10^{-6}$ nm$^{-2}$ K$^{-1}$ and $2.88 \times 10^{-8}$ nm$^{-2}$ K$^{-2}$, respectively and $L_\varphi(0)$ is 56.14 nm. The agreement of fitted curve with observed $L_\varphi(T)$, suggests that temperature dependence of phase coherence length is well explained by equation 2, which confirms the contribution from both e-e and e-ph scattering in magneto-transport of $Bi_{0.85}Sb_{0.15}$ single crystal.

**Conclusion**

In conclusion, the single crystals of $Bi_{0.85}Sb_{0.15}$ were synthesized by self-flux method. Its phase purity and elemental composition were confirmed through XRD and EDAX measurements. In Raman spectra, the active modes were observed for Bi-Bi, Sb-Sb and Bi-Sb bonds. Presence of WAL effect is observed in MR measurements and angle dependent magneto-conductivity study confirms the 2D nature of surface conducting electrons. The low field WAL effect is analysed by HLN model, and the value of prefactor α confirms the conduction from multi-channel. Also, in entire field range the magneto-conductivity is fitted by $HLN + \beta H^2 + \gamma H + c$, confirms the contribution from quantum scattering, cyclotronic MR and bulk carriers in overall conduction. The deviation from Kohler's rule & power law behaviour and temperature dependence of $L_\varphi(T)$ confirms the presence of e-e and e-ph interactions.

**Acknowledgment**

The authors would like to thank Director of National Physical Laboratory (NPL), India, for his keen interest in the present work. Author also thanks to CSIR, India, for research fellowship, and AcSIR-NPL for Ph.D. registration.

**Table 1:** Low field (up to 1.5T) HLN fitted parameters of $Bi_{0.85}Sb_{0.15}$ single crystal

| Temperature (K) | α | $L_\varphi$ (nm) |
|---|---|---|
| 2 | -3.25 | 57.4 |
| 5 | -3.35 | 54.6 |
| 10 | -3.55 | 53.1 |
| 20 | -3.56 | 51.7 |
| 50 | -3.61 | 45.1 |
| 100 | -3.45 | 35.1 |
| 250 | -2.32 | 19.4 |

**Table 2:** Extracted fitted parameters by fitting magneto-conductivity of $Bi_{0.85}Sb_{0.15}$ single crystal using the equation $HLN+\beta H^2+\gamma H+c$.

| Temperature (K) | α | $L_\varphi$ (nm) | β | γ | c |
|---|---|---|---|---|---|
| 2 | -4.05 | 59.7 | -0.01447 | 0.3864 | 0.03831 |
| 5 | -4.03 | 59.6 | -0.01436 | 0.3841 | 0.07363 |
| 10 | -4.31 | 57.7 | -0.01549 | 0.4131 | 0.06351 |
| 20 | -4.31 | 56.6 | -0.01521 | 0.4078 | 0.06527 |
| 50 | -4.46 | 49.1 | -0.01385 | 0.3893 | 0.03781 |
| 100 | -4.12 | 39.1 | -0.00813 | 0.2733 | 0.01707 |
| 250 | -2.49 | 21.5 | -0.00051 | 0.0516 | -0.00431 |

**Figure captions:**

**Fig. 1(a)**: Powder X-ray diffraction spectra of crushed $Bi_{0.85}Sb_{0.15}$ single crystal and solid black curve represents calculated Rietveld refined spectra. **(b)** X-ray diffraction spectra on mechanically cleaved thin flake of $Bi_{0.85}Sb_{0.15}$ single crystal and inset shows the unit cell structure with Bi and Sb atoms as red and black solid symbols, respectively.

**Fig. 2(a)**: Scanning electron microscopy image depicting the layered growth, Energy dispersive X-ray spectroscopy shows elemental composition and mapping reveals the uniform elemental distribution. **(b)** Raman spectra and the corresponding deconvoluted peaks of $Bi_{0.85}Sb_{0.15}$ single crystal measured at room temperature.

**Fig. 3(a)**: Temperature dependent variation of electrical resistivity at different applied magnetic fields. Inset shows the resistance fitted for entire temperature range in zero field. **(b)** Magnetoresistance as a function of transverse magnetic field measured at different temperatures.

**Fig. 4(a):** Shows the magneto-conductivity of $Bi_{0.85}Sb_{0.15}$ single crystal as a function of perpendicular component of magnetic field at 2K. Inset shows the magneto-conductivity vs applied field at different angles at 2K. **(b)** Shows the variation of normalized resistivity as a function of tilt angle between applied field and current direction at 2K and 10T, inset shows the schematic of measurement geometry.

**Fig. 5**: Variation of magneto-conductivity with applied transverse magnetic field at different temperatures, black solid curve shows the fitting with **(a)** HLN equation in low magnetic field (±1.5T) regime and **(b)** HLN+$\beta H^2$+$\gamma H$+c equation in entire field regime.

**Fig. 6(a)**: Kohler's scaling of magnetoresistance as a function of applied magnetic field at different temperatures. **(b)** Variation of phase coherence length with temperature, solid curve shows the fitting using equation 2. Inset shows the deviation from power law behaviour.

Fig. 1(a)

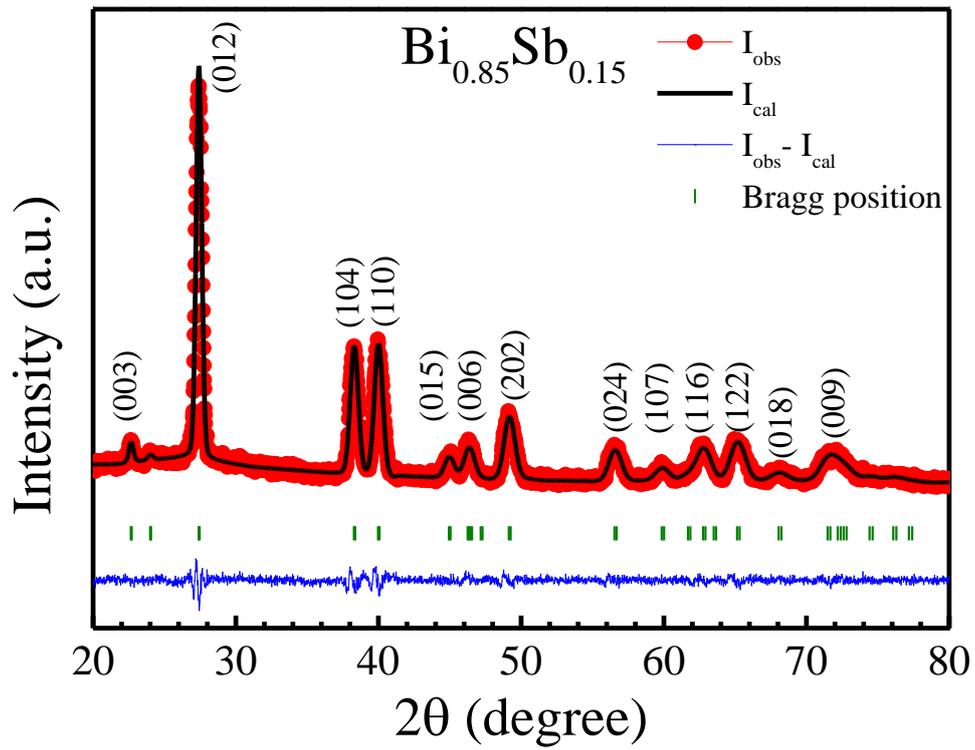

Fig. 1(b)

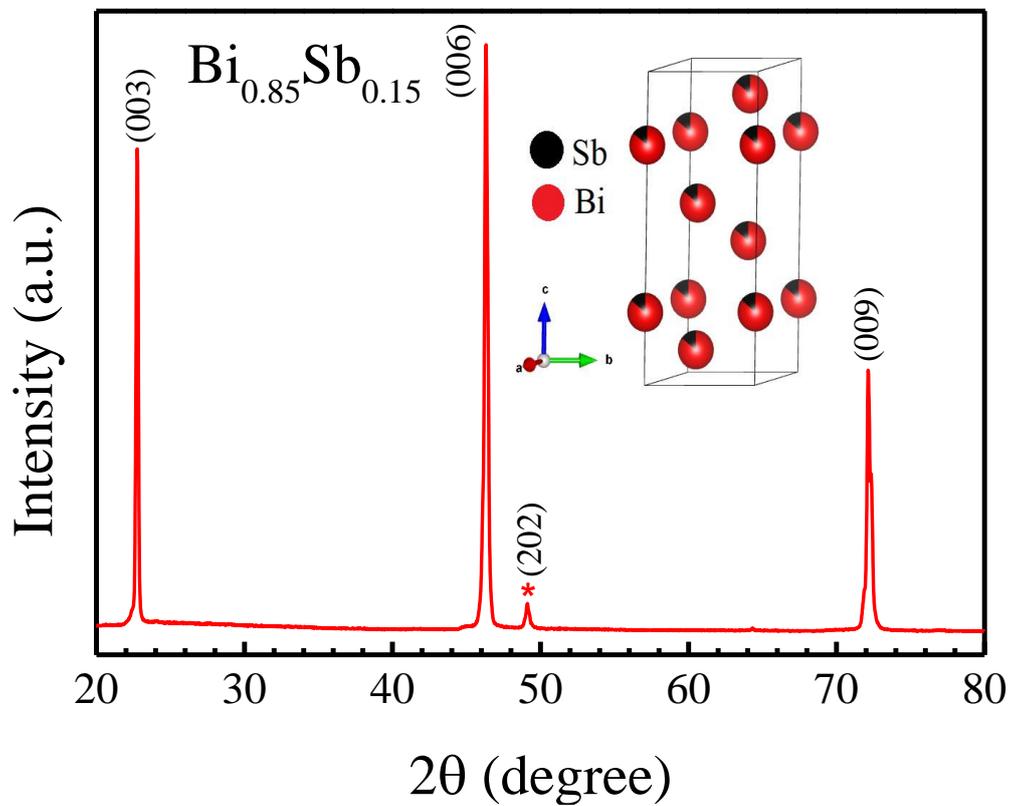

Fig. 2(a)

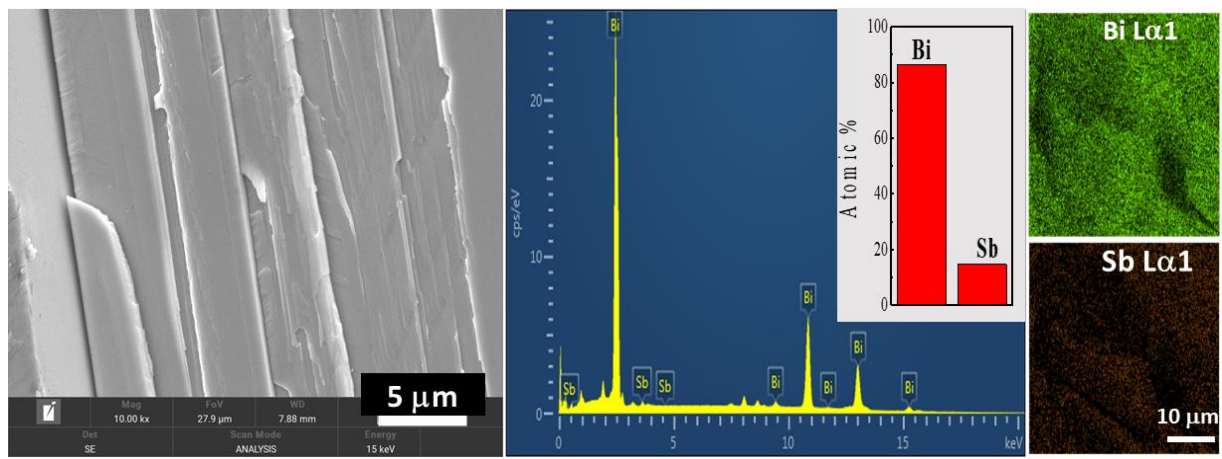

Fig. 2(b)

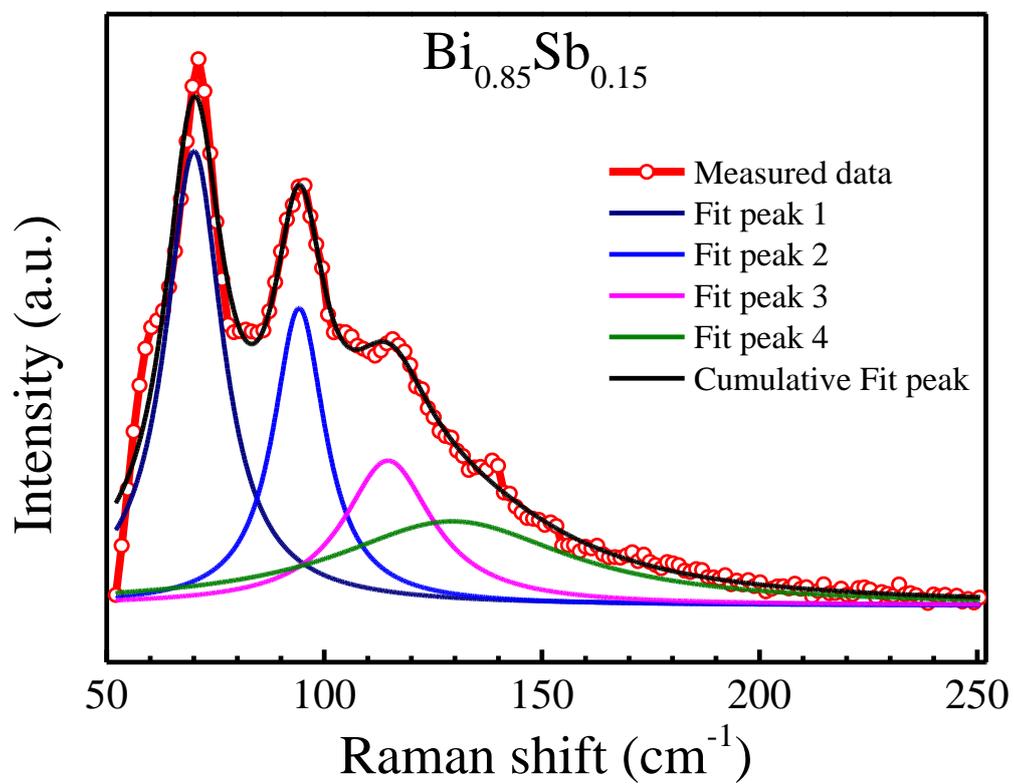

Fig. 3(a)

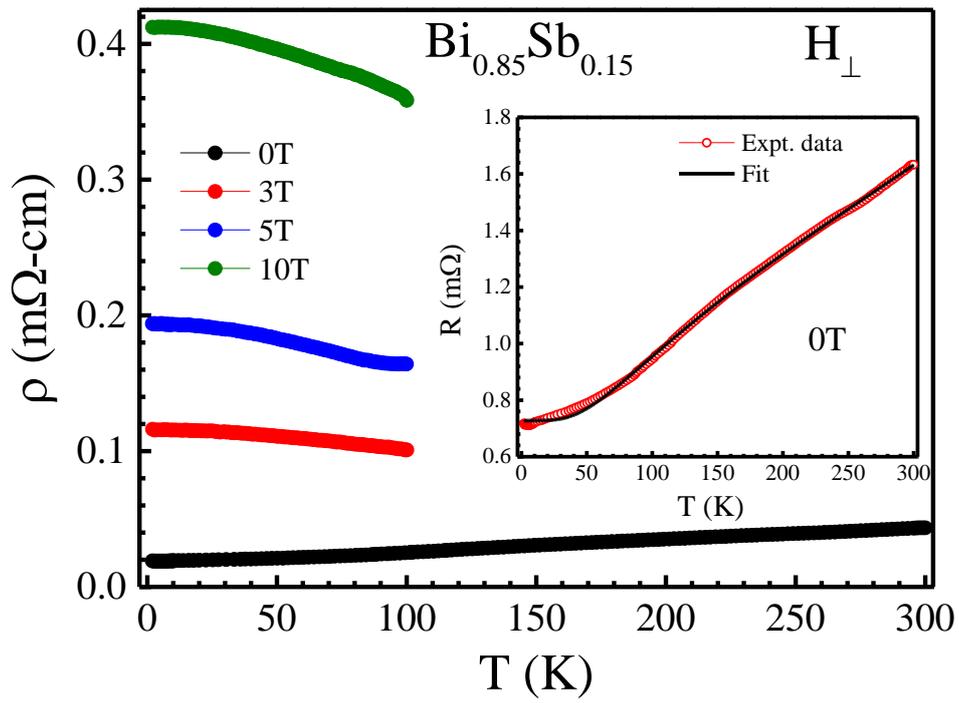

Fig. 3(b)

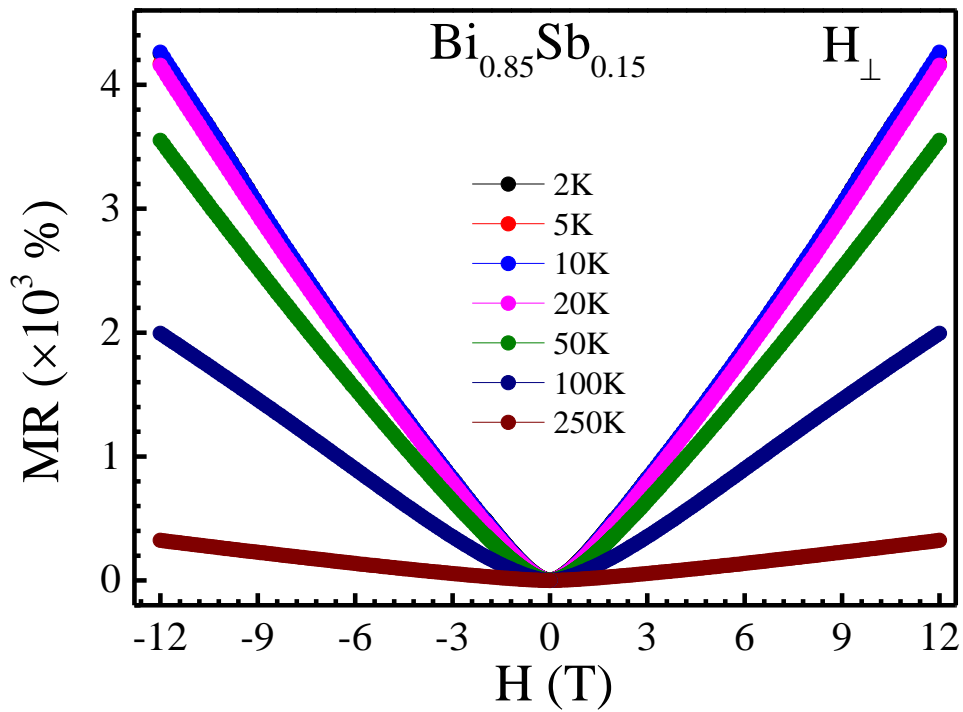

Fig. 4(a)

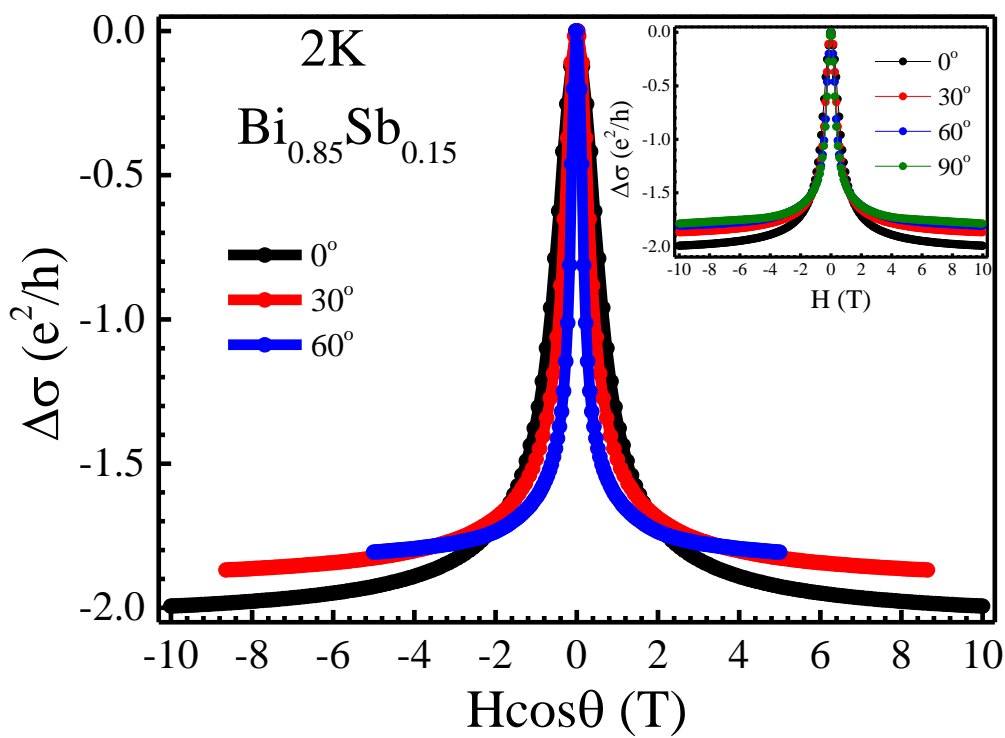

Fig. 4(b)

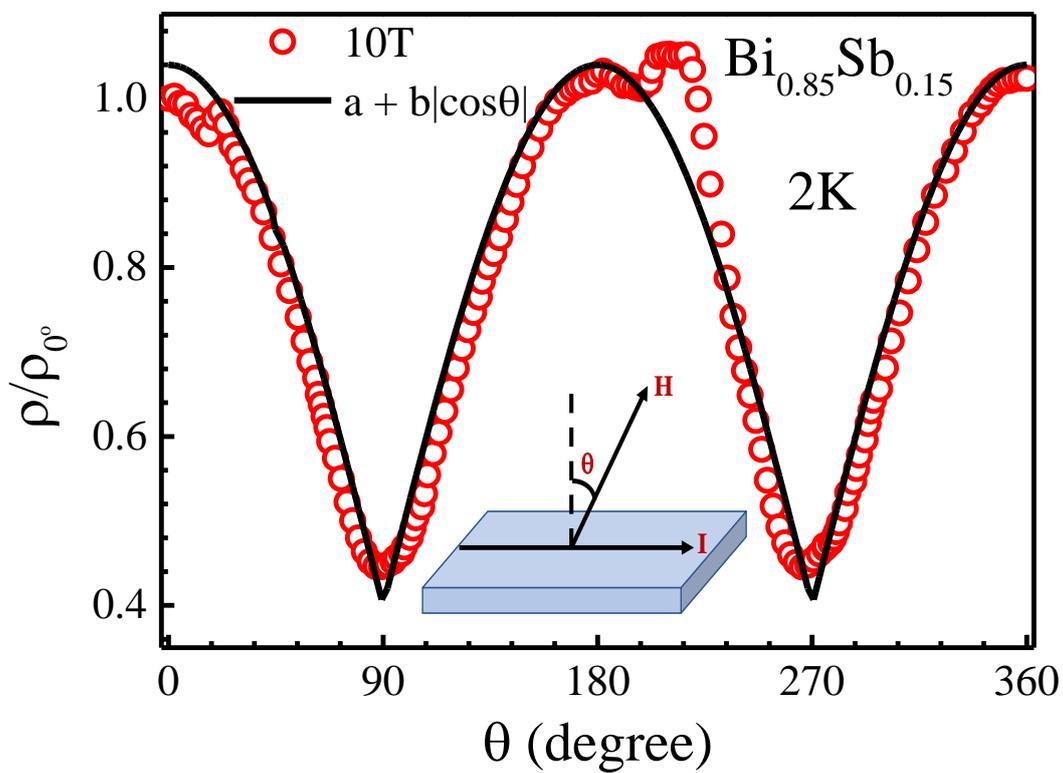

Fig. 5(a)

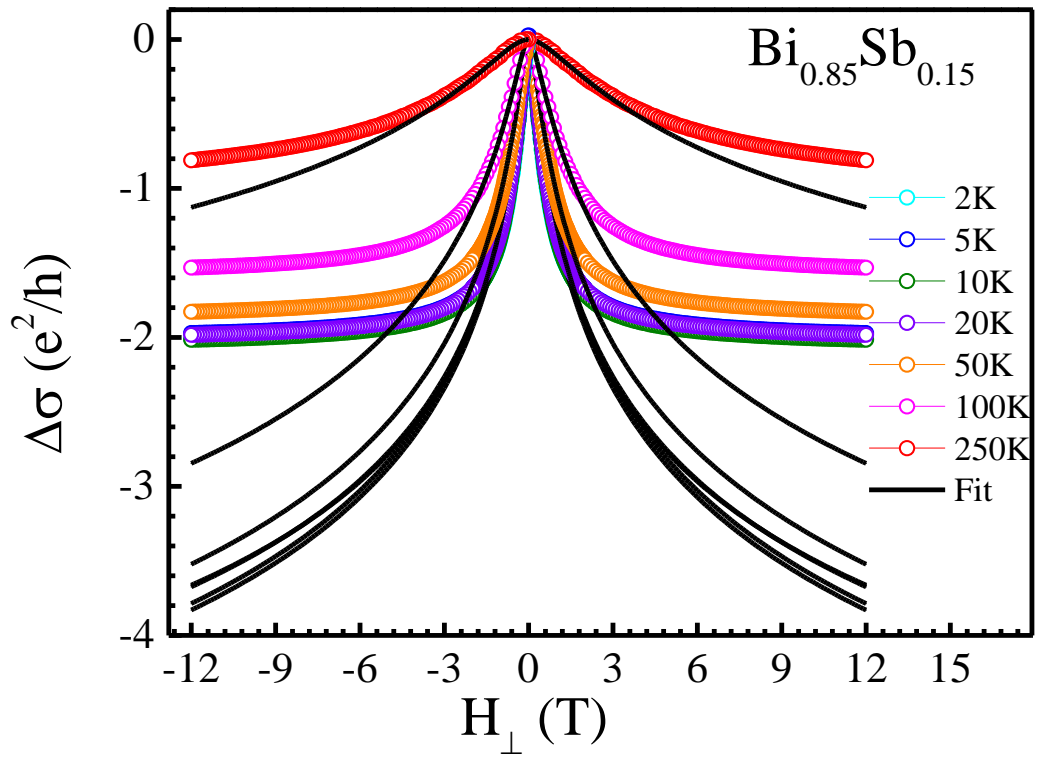

Fig. 5(b):

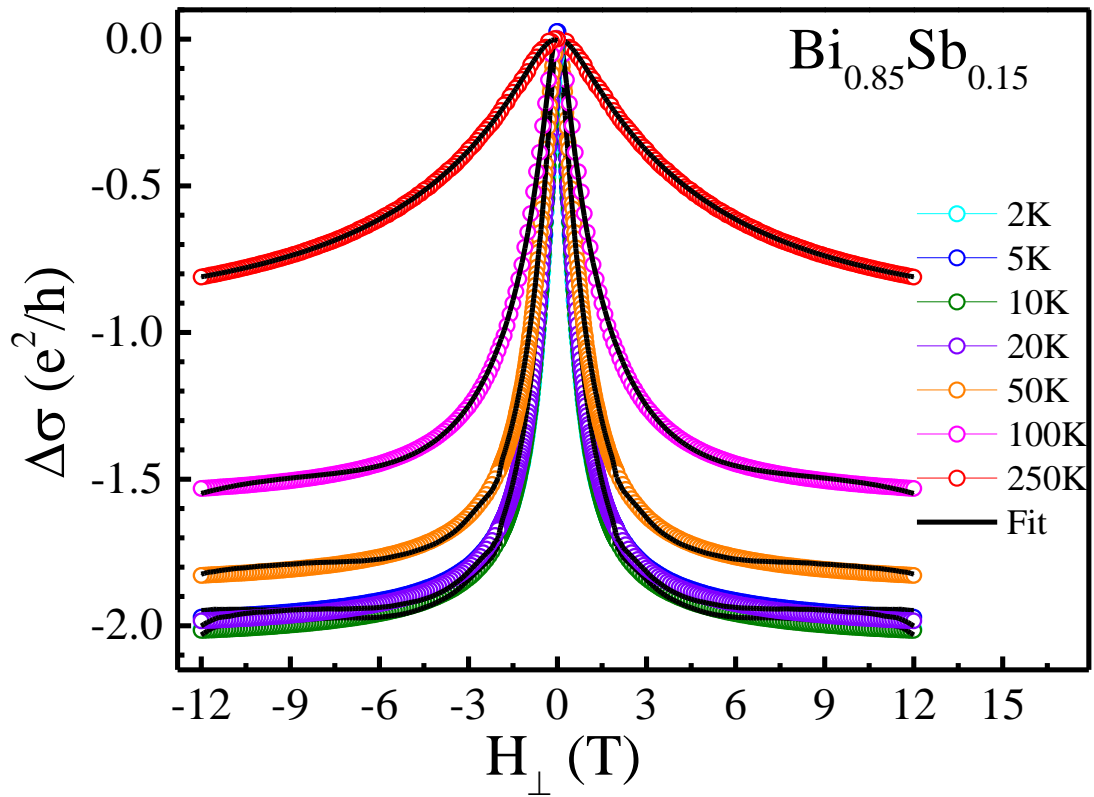

Fig. 6(a)

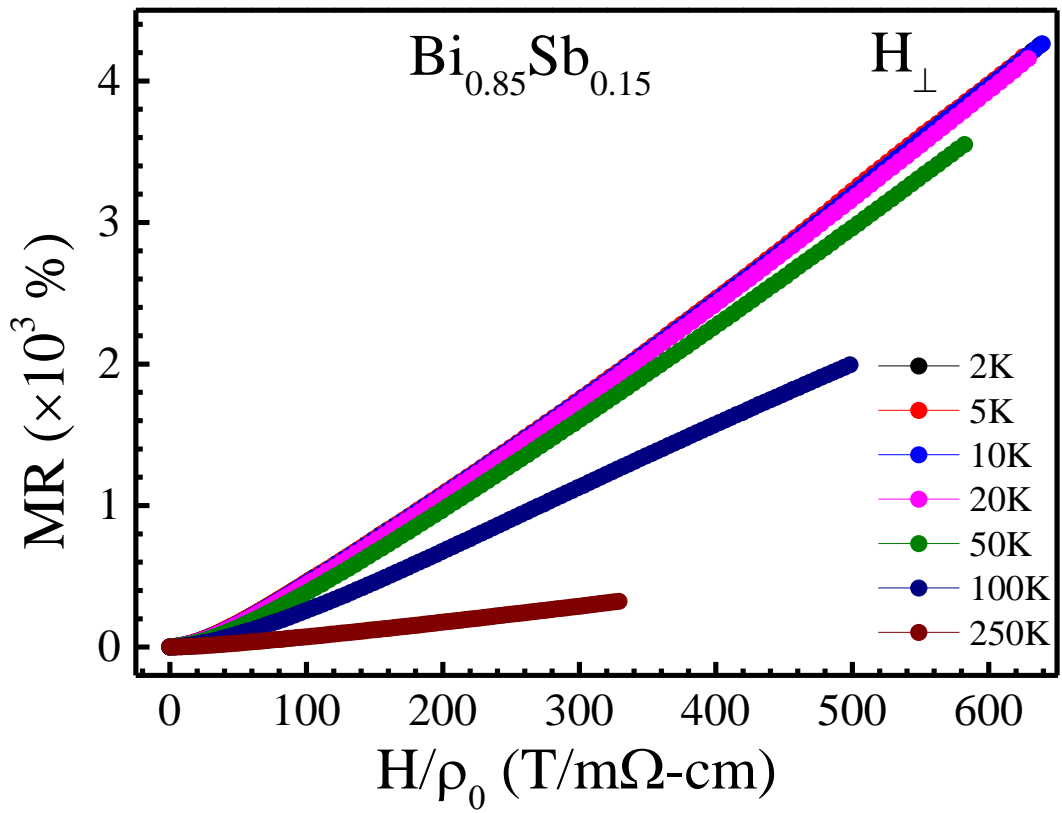

Fig. 6(b)

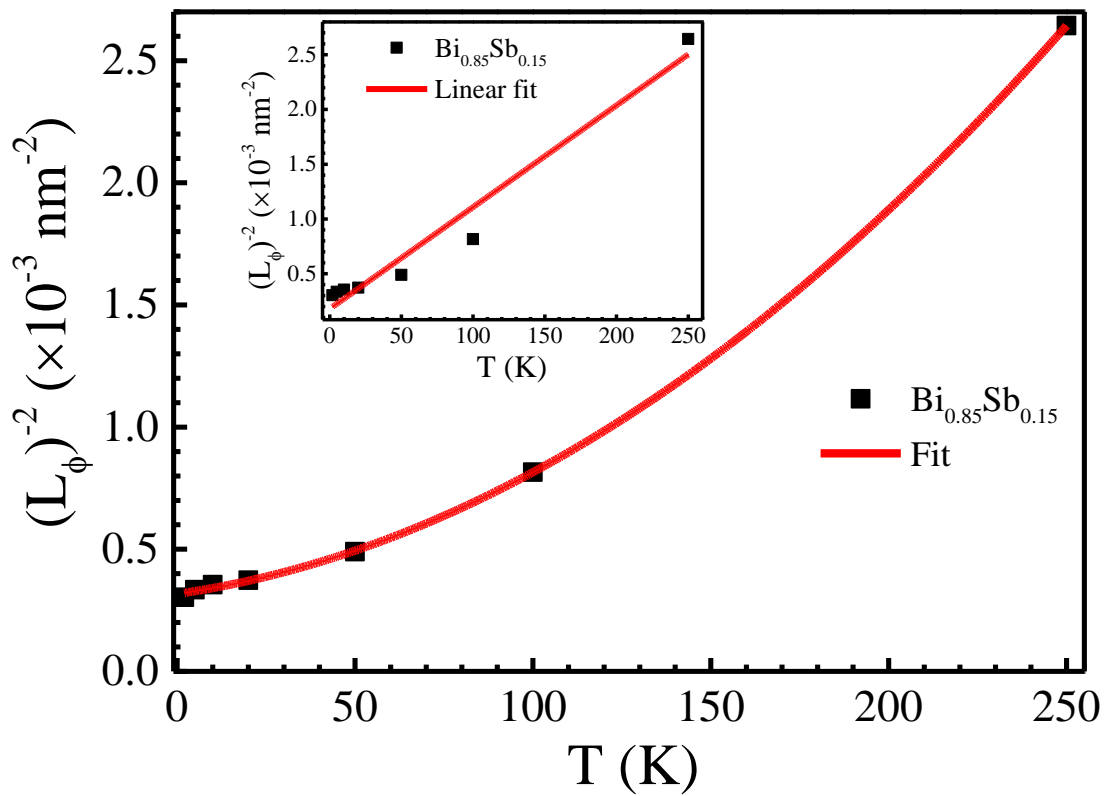